\documentclass[smallcondensed]{svjour3}

\smartqed

\usepackage{graphicx}
\usepackage{amsmath}

\begin{document}

\title{An infinite family of magnetized Morgan-Morgan relativistic thin disks}

\titlerunning{Magnetized Morgan-Morgan relativistic thin disks}     

\author{Antonio C. Guti\'errez-Pi\~neres         \and
        Guillermo A. Gonz\'alez}

\authorrunning{A. C. Guti\'errez-Pi\~neres and G. A. Gonz\'alez} 

\institute{A. C. Guti\'errez-Pi\~neres \at
              Facultad de Ciencias B\'asicas\\
              Universidad Tecnol\'ogica de Bol\'ivar\\
              Cartagena de Indias, Colombia\\
              \email{acgutierrez@unitecnologica.edu.co}   
           \and
           G. A. Gonz\'alez \at
              Escuela de F\'{\i}sica\\
              Universidad Industrial de Santander\\
              A. A. 678, Bucaramanga, Colombia\\
              \email{guillego@uis.edu.co}
}

\date{Received: date / Accepted: date}

\maketitle

\begin{abstract}
Applying the Horsk\'y-Mitskievitch conjecture to the empty space solutions of 
Morgan and Morgan due to the gravitational field of a finite disk, we have 
obtained the corresponding solutions of the Einstein-Maxwell equations. The 
resulting expressions are simply written in terms of oblate spheroidal 
coordinates and the solutions represent fields due to magnetized static thin
disk of finite extension. Now, although the solutions are not asymptotically 
flat, the masses of the disks are finite and the energy-momentum tensor agrees 
with the energy conditions. Furthermore, the magnetic field and the circular 
velocity show an acceptable physical behavior. 
\end{abstract}

\section{Introduction}

The study of axially symmetric solutions of the Einstein and Einstein-Maxwell
field equations corresponding to disklike configurations of matter, apart from
its purely mathematical interest, has a clear astrophysical relevance. Indeed,
thin disks can be used to model accretion disks, galaxies in thermodynamical
equilibrium and the superposition of a black hole and a galaxy. Disk sources for
stationary axially symmetric spacetimes with  magnetic fields are also of
astrophysical importance  mainly in the study of neutron stars, white dwarfs and
galaxy formation.

Exact solutions that have relativistic static thin disks as their sources were 
first studied by Bonnor and Sackfield \cite{BS} and Morgan and Morgan
\cite{MM1,MM2}. Subsequently, several classes of exact solutions corresponding
to static \cite{VOO,LP1,CHGS,LO,LEM,BLK,BLP,GL1,GE,GG-PV} and stationary 
\cite{LP2,BL,PL,GL2} thin disks have been obtained by different authors, and the 
superposition of a static or stationary thin disk with a black hole has been 
considered\cite{LL1,LL2,LL3,SZ1,SEM1,SZ2,SEM2,SEM3,KHZ}. Relativistic disks 
embedded in an expanding FRW universe have been studied in \cite{FIL}, perfect 
fluid disks with halos in \cite{VL1}, and the stability of thin disks models has 
been investigated using a first order perturbation of the energy-momentum tensor 
in \cite{UL1}. On the other hand, thin disks have been discussed as sources for 
Kerr-Newman fields \cite{LBZ,GG1}, magnetostatic axisymmetric fields \cite{LET1}, 
and conformastatic and conformastationary metrics \cite{VL2,KBL,GG-PO}. Also, 
models of electrovacuum static counterrotating dust disks were presented in 
\cite{GG2}, charged perfect fluid disks were studied in \cite{VL3}, and charged 
perfect fluid disks as sources of static and Taub-NUT-type spacetimes in 
\cite{GG3,GG4}.

Now, the thin disks with magnetic fields presented at references
\cite{LBZ,GG1,LET1} were obtained by means of the well known `displace, cut and 
reflect' method in order to introduce a discontinuity at the first derivative of 
one otherwise smooth solution. The result is a solution with a singularity of
the delta function type in all the $z = 0$ hypersurface and so can be 
interpreted as an infinite thi disk. On the other hand, solutions that can be 
interpreted as thin disks of finite extension can be obtained if a proper 
coordinate system is introduced. A coordinate system that adapts naturally to
a finite source and presents the required discontinuous behavior is given by
the oblate spheroidal coordinates. Some examples of finite thin disks from vacuum 
solutions expressed in these coordinates can be found in references \cite{BS,MM1,VOO,LO}, and from electrovacuum solutions in reference \cite{GG-PO}.

According to the above considerations, the purpose of our paper is to present a
new infinite family of exact solutions of the Einstein-Maxwell equations for
axially symmetric spacetimes. The solutions are obtained by the use of  the
Horsk\'y-Mitskievitch conjecture \cite{HM}, which prescribes a quite close
connection between isometries of vacuum spacetimes (seed metrics) and the
electromagnetic four-potential of a generated Einstein-Maxwell fields. We take
the Morgan and Morgan metric disk \cite{MM1} as the seed solution. The generated
solutions describe a family of  magnetized finite thin disks, which is the
magnetized version of the family of relativistic static Morgan and Morgan disks.

The plan of our paper is as follows. First, in Section \ref{sec:einm},  the
Weyl-Lewis-Papapetrou  line element is considered and  Einstein-Maxwell
equations in cylindrical coordinates are introduced. The procedure to obtain
magnetovacuum static, axially symmetric relativistic thin disks without radial
pressure is also summarized in this section. Section \ref{sec:HM} introduces the
Horsk\'y-Mitskievitch generating conjecture used to obtain a general kind of
Weyl-Lewis-Papapetrou spacetime with  magnetic field. In  Section \ref{sec:MMM},
we put forward a solution of the Einstein-Maxwell equations describing an
infinite family of finite static magnetized thin disks by the use of the
Horsk\'y-Mitskievitch (HM) conjecture. For this purpose, the well known Morgan
and Morgan metric disk \cite{MM1} is employed  as seed solution $\Phi_s$.  The
analysis of the physical behavior of solutions is presented in Section
\ref{sec:Beh} where the asymptotic behavior of the solutions is examined.
Then,  we study the behavior of the corresponding energy density, current
density, circular velocity and azimuthal pressure. Finally, in  Section
\ref{sec:conc}, we conclude with a discussion of our results.

\section{Einstein-Maxwell Equations and Thin Disks}\label{sec:einm}

The vacuum Einstein-Maxwell equations, in geometrized units such that 
$c = 8\pi G = \mu _{0} = \epsilon _{0} =  1$, can be written as
\begin{subequations}\begin{eqnarray}
G_{ab} &=& T_{ab},  \label{eq:emep1} \\
{F^{ab}}_{;b} &=& 0, \label{eq:emep2}
\end{eqnarray}\end{subequations}
with the electromagnetic energy-momentum tensor given by
\begin{equation}
T_{ab} = F_{ac} F_b^{ \ c} - \frac{1}{4} g_{ab} F_{cd}
F^{cd} , \label{eq:emtensor}  
\end{equation}
where
\begin{equation*}
F_{ab} =  A_{b,a} - A_{a,b} \label{eq:fab}
\end{equation*}
is the electromagnetic field tensor and $A_a$ is the electromagnetic four
potential. The line element of a static vacuum spacetime can be written in the
Weyl-Lewis-Papapetrou (WLP) form\cite{KSMH}
\begin{eqnarray}
 \mathrm ds^2 = - \mathrm e^{2 \Phi} \mathrm dt^2 \ + \mathrm e^{- 2 \Phi}
 [r^2{\mathrm d}\varphi^2  + \mathrm e^{2 \Lambda} (\mathrm dr^2 + \mathrm
 dz^2)], \label{eq:met0} 
\end{eqnarray}
where ${t,\varphi,r,z}$ are the usual cylindrical coordinates: $-\infty < t,z <
\infty$,\;\;\; $0 \leq r$,\;\;\; $0 \leq \varphi \leq 2\pi$, the metric
functions $\Phi$ and $\Lambda$ depend only on $r$ and $z$. 

The solutions of the Einstein-Maxwell equations corresponding to a disklike
source are even functions of the $z$ coordinate. Therefore, they are continuous
functions everywhere but  their first $z$-derivatives discontinuous at the disk
surface. Consequently, in order to obtain the energy-momentum tensor and the
current density of the source,  the jump across the disk of the first
$z$-derivatives of the metric tensor is expressed as 
\begin{eqnarray}
 b_{ab} =  [{g_{ab,z}}] = 2 {g_{ab,z}}|_{_{z = 0^+}},
\end{eqnarray}
and the jump across the disk of the electromagnetic field tensor is expressed as
\begin{eqnarray}
[F_{za}] = [A_{a,z}] = 2 {A_{a,z}}|_{_{z = 0^+}},
\end{eqnarray}
where the reflection symmetry of the functions with respect to $z = 0$ has
been used.

Then, using the distributional approach \cite{PH,LICH,TAUB} or the junction
conditions on the extrinsic curvature of thin shells \cite{IS1,IS2,POI}, the
Einstein-Maxwell equations yield an energy-momentum tensor as
\begin{equation}
T^{ab} = T_+^{ab} \theta(z) + T_-^{ab} [1 - \theta(z)] + Q^{ab} \delta(z),
\label{eq:emtot}
\end{equation}
and the current density as
\begin{equation}
J^a = I^a \delta(z),   \label{eq:courrent}
\end{equation}
where $\theta(z)$ and $\delta (z)$ are, respectively, the Heaviside and Dirac
distributions with support on $z = 0$. Here $T_\pm^{ab}$ are the electromagnetic
energy-momentum tensors as they are defined by (\ref{eq:emtensor}) in the $z
\geq 0$ and
$z \leq 0$ regions, respectively, whereas 
\begin{eqnarray}
Q^a_b = \frac{1}{2}\{b^{az}\delta^z_b - b^{zz}\delta^a_b + g^{az}b^z_b -
g^{zz}b^a_b 
+ b^c_c (g^{zz}\delta^a_b - g^{az}\delta^z_b)\},
\end{eqnarray}
gives the part of the energy-momentum tensor corresponding to the disk source,
and  
\begin{equation}
I^a  =[F^{az}] \label{eq:currenti}
\end{equation}
is the contribution of the disk source to the current density. The surface
energy-momentum tensor, $S_{ab}$, and the surface current density,
$j^a$, of the disk, can be obtained through the relations
\begin{eqnarray}
S_{ab} \ = \ \int Q_{ab} \ \delta (z) \ ds_n \ = \ e^{ \Lambda - \Phi} \
Q_{ab} ,
\end{eqnarray}
and
\begin{eqnarray}
j_a \ =  \ \int I_a \ \delta (z) \ ds_n \ = \  e^{\Lambda - \Phi} I_a ,
\label{eq:currentj}\end{eqnarray}
where $ds_n = \sqrt{g_{zz}} \ dz$ is the physical measurement of length in
the normal direction  of the disk. 

Now, we choose the magnetic potential as 
\begin{equation}
A_a = (0, A(r,z), 0, 0) \label{eq:empot},
\end{equation}
so that we have a pure axially symmetric magnetic field
\begin{eqnarray}
F_{\varphi z} = -A_{,z},\;\;\;\;\; F_{r\varphi} = A_{,r}.
\end{eqnarray}
Thus, for the metric (\ref{eq:met0}), the only non-zero components of $S^a_b$
and  $j_a$ are
\begin{eqnarray}
S^0_0 &=& \ 2 e^{\Phi - \Lambda} \left\{ \Lambda,_z - \ 2 \Phi,_z\right\},
\nonumber\\
S^1_1 &=& \ 2 e^{\Phi - \Lambda} \Lambda,_z ,\label{eq:emt22}\\
j_{\varphi} &=& \  -2e^{\Phi - \Lambda} A_{,z},\nonumber
\end{eqnarray}
where all the quantities are evaluated at $z = 0^+$. 
	
With the orthonormal tetrad  
\begin{eqnarray}
e_{(b)}^{\;\;a}=\{V^a,W^a,X^a,Y^a\} \label{eq:tet},
\end{eqnarray}
 where
\begin{subequations}
\begin{eqnarray}
V^a &=& e^{-\Phi} (1, 0, 0, 0 ),\\ 
W^a &=& \frac{e^{\Phi}}{r} (0, 1, 0, 0 ),\\ 
X^a &=& e^{\Phi - \Lambda} (0, 0, 1, 0 ),\\
Y^a &=& e^{\Phi - \Lambda} (0, 0, 0, 1 ),
\end{eqnarray}
\end{subequations}
we can write the surface energy-momentum tensor of the disk in the canonical
form as
\begin{eqnarray}\label{eq:emt2}
S^{ab} = \epsilon V^a V^b + p W^aW^b.
\end{eqnarray}
Here $\epsilon$  and $p$ are the energy density and the azimuthal pressure of
the disk, respectively. We also consider the mass density on the surface of the
disk, defined as
\begin{eqnarray}
\mu = \epsilon + p.
\end{eqnarray}
In the same way, the current density on the disk can be written as
\begin{eqnarray}j^a = j W^a\label{eq:currentet}.
\end{eqnarray}
As we can see the charge density  on the disk surface is identically equal to
zero. This can be checked by using Eqs. (\ref{eq:currenti}),
(\ref{eq:currentj}), (\ref{eq:empot}) and (\ref{eq:currentet}).  The
electromagnetic field is of a magnetic type, which can be demonstrated by the
electromagnetic invariant
\begin{eqnarray}
F_{ab}F^{ab}= \frac{2e^{(2\Phi - \Lambda)}}{r^2}[A^2_{,r} + A^2_{,z}] \geq 0.
\label{eq:hp2}\end{eqnarray}

We consider now, on the basis of Refs. \cite{GE} and \cite{GG4}, the 
possibility that the energy-momentum tensor $S^{ab}$ and the current density
$j^{a}$ can be interpreted as the superposition of two counterrotating fluids.
In order to do this, we cast
\begin{subequations}\begin{eqnarray}
S^{ab} &=& \epsilon_+U^a_+U^b_+ + \epsilon_-U^a_-U^b_- ,\\
&  \nonumber\\
j^a &=& \sigma_+U^a_+ + \sigma_-U^a_- ,
\end{eqnarray}
\end{subequations}
where
\begin{subequations}\begin{eqnarray}
\epsilon_+ &=& \epsilon_- = (\epsilon - p) /2,\\
& \nonumber\\
\sigma_+ &=& - \sigma_- = \frac{je^{\Phi}}{2r}\sqrt{\frac{\epsilon}{p} - 1},
\end{eqnarray}\end{subequations}
are the energy densities and charge densities of the two counterrotating fluids.
The counterrotating  velocity vectors are given by 
\begin{eqnarray}
U^a_{\pm}=\frac{V^a \pm UW^a}{\sqrt{1 - U^2}},
\end{eqnarray}
where
\begin{eqnarray}
U^2= \frac{p}{\epsilon}  \leq 1
\end{eqnarray} 
is the counterrotating tangential velocity. Therefore, we have two
counterrotating charged fluids with equal energy densities and equal but
opposite charge densities.

\section{The Horsk\'y-Mitskievitch conjecture and the WLP metric}
\label{sec:HM}
\subsection{The seed WLP metric}

The WLP line element of a static vacuum spacetime can be written as \cite{KSMH}
\begin{eqnarray}
ds^2 = - \ e^{2 \Phi_s} dt^2 + 
e^{- 2 \Phi_s}[r^2  d\varphi^2 + e^{2\Lambda_s} (dr^2 +dz^2)], \label{eq:met}
\end{eqnarray} 
where $\{t,\varphi,r,z\}$ are the usual cylindrical coordinates: $-\infty < t,z
<
\infty$, $0 \leq r$, $0 \leq \varphi \leq 2\pi$. The metric functions $\Phi_s$
and $\Lambda_s$ depend on $r$ and $z$. This metric admits two Killing
vectors:
\begin{subequations}
\begin{eqnarray}
\boldsymbol{\xi_{t}} &=& (1,0,0,0),\\
\boldsymbol{\xi_{\varphi}} &=& (0,1,0,0),
\end{eqnarray}
\end{subequations}
which, in the vectorial space $\partial_i$ are,
\begin{subequations}
\begin{eqnarray}
\boldsymbol{\xi_{t}} &=& \delta^j_t\partial_j,\\
\boldsymbol{\xi_{\varphi}} &=& \delta^j_\varphi\partial_j,
\end{eqnarray}
\end{subequations}
or, in terms of their components are
\begin{subequations}
\begin{eqnarray}
\xi^j_{t} &=& \delta^j_t,\\
\xi^j_{\varphi} &=& \delta^j_\varphi,
\end{eqnarray}
\end{subequations}
where $\delta^a_b$ is the usual Kronecker delta tensor.

In terms of the one-forms $\boldsymbol{dt}$ and $\boldsymbol{d\varphi}$  we
have 
\begin{subequations}
\begin{eqnarray}
\boldsymbol{\xi_{t}} &=& -e^{2\Phi_s}\boldsymbol{dt}\label{eq:kil1},\\
\boldsymbol{\xi_\varphi} &=& r^2e^{-2\Phi_s}\boldsymbol{d\varphi}\label{eq:kil2},
\end{eqnarray}
\end{subequations}
where all the orthonormal bases employed  are always chosen as a generalization
of the set
\begin{subequations}
\begin{eqnarray}
\boldsymbol{\omega^{_{(0)}}} &=& e^{\Phi_s}\boldsymbol{dt},\\
\boldsymbol{\omega^{_{(1)}}} &=& re^{-\Phi_s}\boldsymbol{d\varphi},\\
\boldsymbol{\omega^{_{(2)}}} &=& re^{\Lambda_s-\Phi_s}\boldsymbol{dr},\\
\boldsymbol{\omega^{_{(3)}}} &=& re^{\Lambda_s-\Phi_s}\boldsymbol{dz},
\end{eqnarray}
\end{subequations}
the simplest tetrad  that can be used for the Weyl solution. 

\subsection{The WLP solution with a magnetic field}
The present section is devoted to the derivation of solutions of the 
Einstein-Maxwell equations  via HM conjecture. Following the method used in
\cite{RNH}, we modify the line element (\ref{eq:met}) to the form
\begin{eqnarray}
 ds^2 = - f(r,z)^2\ e^{2 \Phi_s} dt^2 + \frac{e^{- 2 \Phi_s}r^2}{f(r,z)^2} 
d\varphi^2 + f(r,z)^2e^{2(\Lambda_s -\Phi_s)} (dr^2 +dz^2), \label{eq:met2} 
\end{eqnarray} 
that admits two Killing vectors
 \begin{subequations}
\begin{eqnarray}
\boldsymbol{\xi_{t}} &=& -f(r,z)^2e^{2\Phi_s}\boldsymbol{dt},\label{eq:kil1a}\\
\boldsymbol{\xi_{\varphi}}&=&
\frac{r^2e^{-2\Phi_s}}{f(r,z)^2}\boldsymbol{d\varphi}\label{eq:kil2a},
\end{eqnarray}
\end{subequations}
with $f(r,z)$  an arbitrary function. 

The Killing vectors $\boldsymbol{\xi}$ and the electromagnetic four-potential
$\boldsymbol{A}$ satisfy  
\begin{eqnarray}
\boldsymbol{^\star d^\star d\xi} = 0,\;\;\;\; 
\boldsymbol{^\star d^\star d A} = 0,
\end{eqnarray}
where the $\boldsymbol{^{\star}}$  is the usual Hodge (star) operation. Then, the 
Killing vector $\boldsymbol{\xi_{\varphi}}$ induces the four-potential
\begin{eqnarray}
\boldsymbol{A} = qf(r,z)\boldsymbol{\xi_{\varphi}},
\end{eqnarray}
or,
\begin{eqnarray}
\boldsymbol{A} = \frac{qr^2}{fe^{2\Phi_s}}\boldsymbol{d\varphi}.
\label{eq:magpot}\end{eqnarray}
It can be verified through standard calculations that the sourceless
Einstein-Maxwell equations are fulfilled if 
\begin{eqnarray}
f(r,z)= 1 + c_1f_1(r,z),
\end{eqnarray} 
where the function $f_1$ must be a solution of the differential equation
\begin{eqnarray}
G^{(a)}_{\;\;\;\;(a)} = - R = 0,
\end{eqnarray}
given the fact, for electrovacuum spacetimes, the
Einstein tensor,  $G_{(a)(b)}$, is traceless. As in \cite{RNH} we can
choose $c_1=q^2$ and $f_1=r^2e^{-2\Phi_s}$, and then we have the four-potential
\begin{eqnarray}
{\boldsymbol A}= \frac{qr^2}{q^2r^2  + e^{2\Phi_s}}\boldsymbol{d\varphi},
\end{eqnarray}
the $q$ parameter characterizing the strenght of the electromagnetic field. 
Under the conditions described above the metric set by (\ref{eq:met2}) is
reduced to
\begin{eqnarray}
ds^2 = - \ e^{2 \Phi} dt^2 + 
e^{- 2 \Phi}[r^2  d\varphi^2 + e^{2\Lambda} (dr^2 +dz^2)], \label{eq:met3}
\end{eqnarray} 
where,\;\;$fe^{\Phi_s} = e^{\Phi}$ and $ f^2e^{\Lambda_s} = e^{\Lambda}$.

Using  (\ref{eq:emt22}),  (\ref{eq:tet}) and (\ref{eq:emt2}) we find out the
surface energy density
\begin{eqnarray}
\epsilon=\frac
{4E^2(r)e^{3\Phi_s - \Lambda_s}(1 - r\Phi_{s,r})\Phi_{s,z}}
{(e^{2\Phi} + q^2r^2)^3},\label{eq:energy}
\end{eqnarray}
the azimuthal pressure
\begin{eqnarray}
p=\frac{4rE(r)e^{3\Phi_s - \Lambda_s}[2q^2r + E(r)\Phi_{s,r}]\Phi_{s,z}}
{(e^{2\Phi_s} + q^2r^2)^3}\label{eq:pressure},
\end{eqnarray}
the surface mass density
\begin{eqnarray}
\mu=\frac{4E(r)e^{3\Phi_s - \Lambda_s}\Phi_{s,z}}
{(e^{2\Phi_s} + q^2r^2)^2}\label{eq:mass},
\end{eqnarray}
the surface current density
\begin{eqnarray}
j=\frac{4qre^{4\Phi_s - \Lambda_s}\Phi_{s,z}}
{(e^{2\Phi_s} + q^2r^2)^2}\label{eq:current},
\end{eqnarray}
and the circular velocity
\begin{eqnarray}
U^2= U_s^2\left[ 1 + \frac{2q^2r}{E(r)\Phi_{s,r}} \right]
\label{eq:velocity},
\end{eqnarray}
where
$E(r)= e^{2\Phi_s} - q^2r^2$, 
and
\begin{eqnarray}
U_s^2= \frac{r\Phi_{s,r}}{1- r\Phi_{s,r}}
\end{eqnarray}
is the circular velocity of the unmagnetized source, i.e. $q=0$. 

As we can see from (\ref{eq:kil1}), (\ref{eq:kil2}), (\ref{eq:kil1a}) and
(\ref{eq:kil2a}), the method is a solution generating technique that changes
the norm  of the two Killing vectors without adding twist. Besides, the possible
connection of the HM conjecture with inner symmetries of Einstein-Maxwell
equations  would connect the conjecture  with a set of generating
methods elaborated by Ernst and other authors for axisymmetric fields \cite{RH}.

\section{Magnetized Morgan-Morgan disks}\label{sec:MMM}

Let's restrict the previous general model to obtain a solution of the
Einstein-Maxwell equations describing an infinite family of finite static
magnetized thin disks. For this purpose, we use, as a seed solution
$(\Phi_s,\;\Lambda_s)$, the well known Morgan and Morgan metric disk  $(\phi ,\;
\lambda)$ \cite{MM1}, 
\begin{eqnarray*}
 \phi(x,y)=-\sum_{n=0}^{\infty}
C_{2n}q_{2n}(x)P_{2n}(y), 
\end{eqnarray*}  
where  $C_{2n}$ are arbitrary constants that must be properly specified so that
a  particular solution may be set. $P_{n}(y)$ are the usual Legendre polynomials
and $q_n(x)=i^{n+1}Q_n(ix)$.\; $Q_n(z)$ being the Legendre functions of second
kind (see \cite{AW} and, for the Legendre functions of imaginary argument,
\cite{MF} pag. 1328). The  $x$ and $y$ are the oblate spheroidal
coordinates related with the cylindrical coordinates by the relations \cite{MF} 
\begin{subequations}\begin{eqnarray} 
r^{2} &=& a^{2}(1 + x^{2})(1 - y^{2}), \label{eq:ciloblatas1} \\
z &=& a xy, \label{eq:ciloblatas2}
\end{eqnarray}\end{subequations}
where $0 \leq x < \infty$ and $-1 \leq y < 1$. The disk has coordinates
$x = 0$, $0\leq y^2<1$ and, when the disk is crossed, the sign of $y$
changes, but not its absolute value.

We use  the constants $C_{2n}$ as they have been determined by Gonz\'alez and Reina
in \cite{GR}
\begin{eqnarray*}
C_{2n}= \frac{mG}{2a}\left[\frac{\pi^{1/2}(4n +1)(2l +1)!}
{2^{2l}(2n +1)(l-n)!\Gamma(l+n + \frac{3}{2})q_{2n+1}(0)}\right],
\label{recu}
\end{eqnarray*}
for $n \leq l$ and $C_{2n}= 0$ for $n > l$. Consequently, these solutions
correspond to the magnetized version of the well known Morgan-Morgan
disk \cite{MM1}. For instance, the first three members of the family of the seed 
metric functions are given by 
\begin{subequations}
\begin{eqnarray}
 \phi_1(x,y) &=& - \frac{mG}{a} [ \cot^{-1}x  + A (3y^{2} - 1)],
\label{eq:4.22}   \\
 \phi_2(x,y) &=& - \frac{mG}{a} [ \cot^{-1} x + \frac{10 A}{7}
(3y^{2} - 1)  + \ B ( 35 y^{4} - 30 y^{2} + 3)], \label{eq:4.23} \\
 \phi_3(x,y) &=& - \frac{mG}{a} [ \cot^{-1} x + \frac{10 A}{6} (3
y^{2} - 1)   + \label{eq:4.24} \\ &&\frac{21 B}{11} (35 y^{4} - 30 y^{2} + 3)
  + \ C (231 y^{6} - 315 y^{4} + 105 y^{2} - 5) ],
\nonumber
\end{eqnarray}\end{subequations}
where
\begin{subequations}
\begin{eqnarray*}
&& A = \frac{1}{4} [(3x^{2} + 1) \cot^{-1} x - 3 x ],
\\
&& B = \frac{3}{448} [ (35 x^{4} + 30 x^{2} + 3) \cot^{-1} x
- 35 x^{3} - \frac{55}{3} x ], \\
&& C = \frac{5}{8448} [ (231 x^{6} + 315 x^{4} + 105 x^{2} +
5) \cot^{-1} x  - 231 x^{5} - 238 x^{3} - \frac{231}{5} x ],
\end{eqnarray*}\end{subequations}
which similar, but more involved, expressions for greater values of $l$.

On the other hand,  according to (\ref{eq:energy}), the energy density of
the magnetized disk can be written as
\begin{eqnarray}
\epsilon_n=\frac{4E^2e^{3\phi_n - \lambda_n}
[y + (1 -y^2)\phi_{n,y}]\phi_{n,x}}
{ay^2[e^{2\phi_n} + q^2a^2(1 -y^2)]^3},
\end{eqnarray}
with
\begin{eqnarray*}
E(y)= e^{2\phi_n} -q^2a^2(1 -y^2),
\end{eqnarray*}
while, according to (\ref{eq:pressure}) and (\ref{eq:mass})
the pressure, $p$, and mass density, $\mu$, in the surface of the disk are
\begin{eqnarray}
p_n=\frac{4E(1 -y^2)e^{3\phi_n - \lambda_n}\phi_{n,x}
[2q^2a^2y - \phi_{n,y}E]}
       {ay^2[e^{2\phi_n} + q^2a^2(1 -y^2)]^3},
\end{eqnarray}
and
\begin{eqnarray}
\mu_n = \frac{4Ee^{3\phi_n - \lambda_n}\phi_{n,x}}
     {ay[e^{2\phi_n} + q^2a^2(1 -y^2)]^2},
\end{eqnarray}
respectively, and,  according to  (\ref{eq:current}) and
(\ref{eq:velocity}),  we have  the following  surface current density and
circular
velocity
\begin{eqnarray}
j= \frac{4q(1 -y^2)^{1/2}e^{4\phi_n - \lambda_n}\phi_{n,x}}
     {y[e^{2\phi_n} + q^2a^2(1 -y^2)]^2},
\end{eqnarray}
and
\begin{eqnarray}
U^2_n= U_s^2\left[1 - \frac{2q^2a^2y}{E\phi_{n,y}}\right],
\end{eqnarray}
 respectively, where 
\begin{eqnarray*}
U_s^2=-\frac{(1 - y^2)\phi_{n,y}}{y + (1 -y^2)\phi_{n,y}},
\end{eqnarray*}
is the circular velocity of the unmagnetized source, i.e. $q=0$. The non zero 
components of the magnetic field on the surface of the disk, $B_z= - 
A_{\varphi,r}$ and $B_r = A_{\varphi,z}$, are 
\begin{eqnarray}
B_z= -\frac{2qa(1 -y^2)^{1/2}e^{2\phi_n}
[y + (1-y^2)\phi_{n,y}]}
{y[e^{2\phi_n} + q^2a^2(1 - y^2)]^2},
\end{eqnarray}
and
\begin{eqnarray}
B_r =  -\frac{2qa(1 - y^2)e^{2\phi_n}\phi_{n,x}}
{y[e^{2\phi_n} + q^2a^2(1 - y^2)]^2},
\end{eqnarray}
where we have used the magnetic potential of (\ref{eq:magpot}), and
all the quantities are evaluated on the disk.

\section{The behavior of the solutions \label{sec:Beh}}
\begin{figure}
\begin{center}
\includegraphics[width=3.2in]{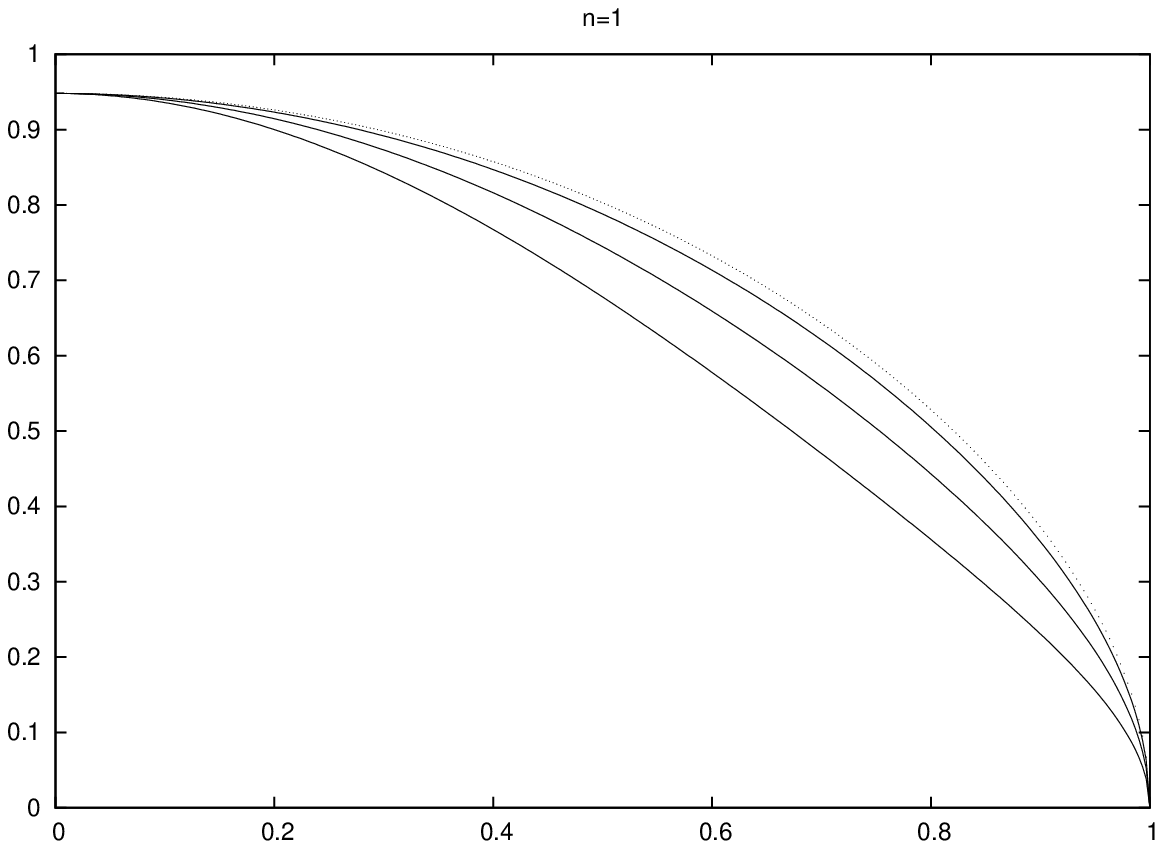}
\includegraphics[width=3.2in]{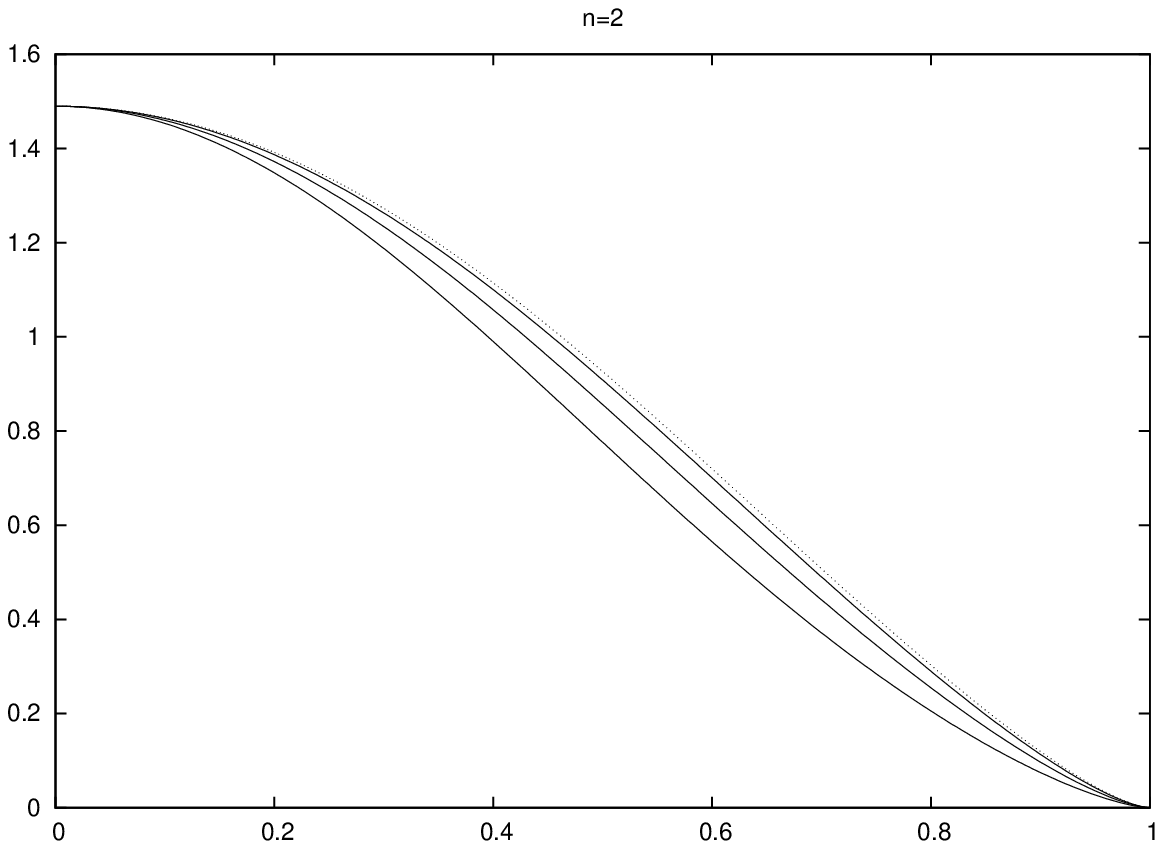}
\includegraphics[width=3.2in]{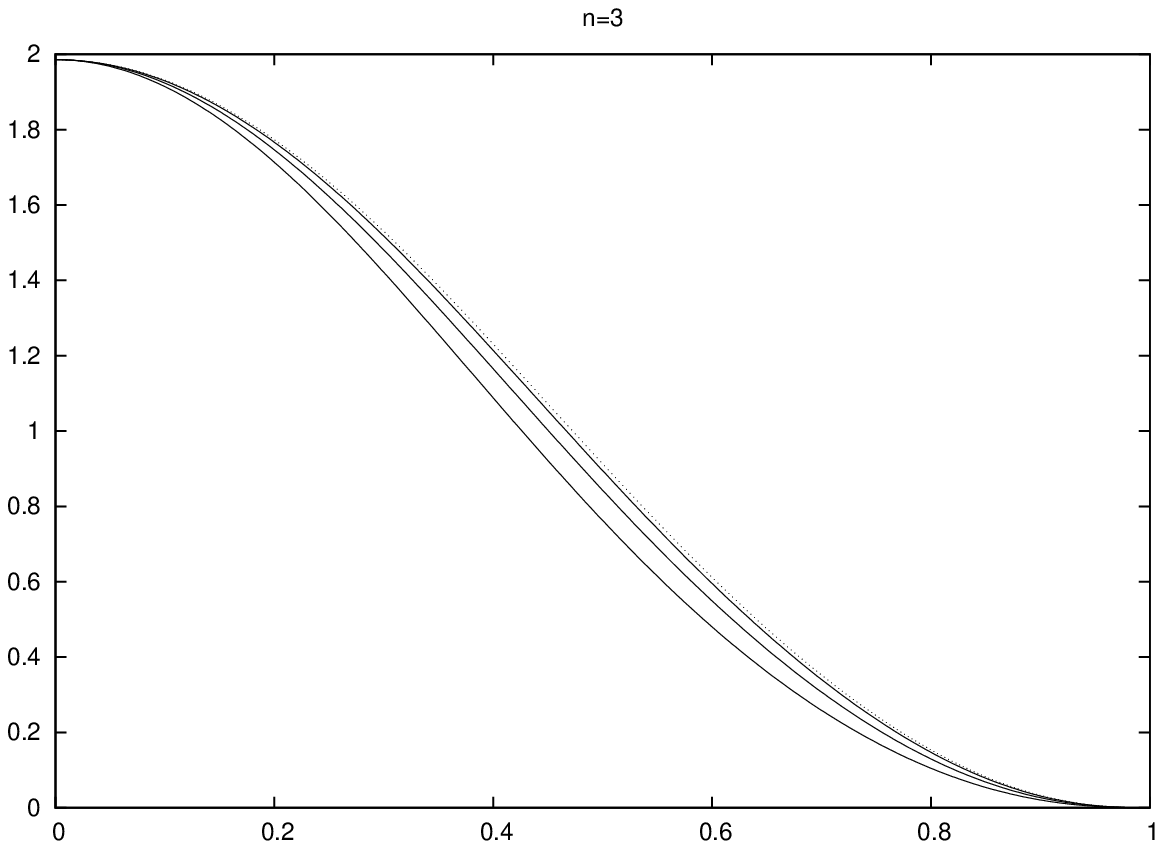}
\caption{\label{fig:energy} Dimensionless energy density ${\tilde \epsilon}_n$
as a function of ${\tilde r}$ for the first three  disk models  with $n=1,2,3$.
In each case, we plot ${\tilde \epsilon}_n(\tilde r)$ for $0 \leq \tilde r \leq
1$ with $m = 0.10$ for different values of the parameter $\tilde q$. First, we
take $\tilde q =0.30$, the bottom curve in each plot, and then $0.20,\;0.10$ and
$\tilde q = 0$ (curve with dots), the vacuum case.}
\end{center}
\end{figure}

In order to elucidate the behavior of the different particular models, firstly
we introduce the dimensionless energy surface density on the disks, defined 
for $0 \leq \tilde r \leq 1$ as
$
\epsilon_n(r) = a \tilde \epsilon_n(\tilde r),
$
where  the dimensionless radial variable  $r = a \tilde r$ has been introduced.
In figure \ref{fig:energy}  the dimensionless surface energy densities
${\tilde \epsilon}_n$  for the models corresponding  to $n=1,\;2$ and $3$ are
depicted. In each case, ${\tilde \epsilon}_n(\tilde r)$ for $0 \leq \tilde r
\leq 1$ with $m = 0.10$ for different values of the parameter $\tilde q = aq$
are shown. First, we set $\tilde q =0.30$, the bottom curve in each graphic, and
then $ 0.20,\;0.10$ and $\tilde q = 0$ (dotted curve), the vacuum case. In all
the cases, it can be seen that the energy density  is everywhere positive
and that it vanishes at the edge of the disk. The disks with higher values of
$n$ and $\tilde q$ show an energy distribution that is more concentrated in the
center and lower at the edge. It can be  observed that the energy density in the
central region of the disk decreases with the presence of a magnetic field. 

\begin{figure}
\begin{center}
\includegraphics[width=3.2in]{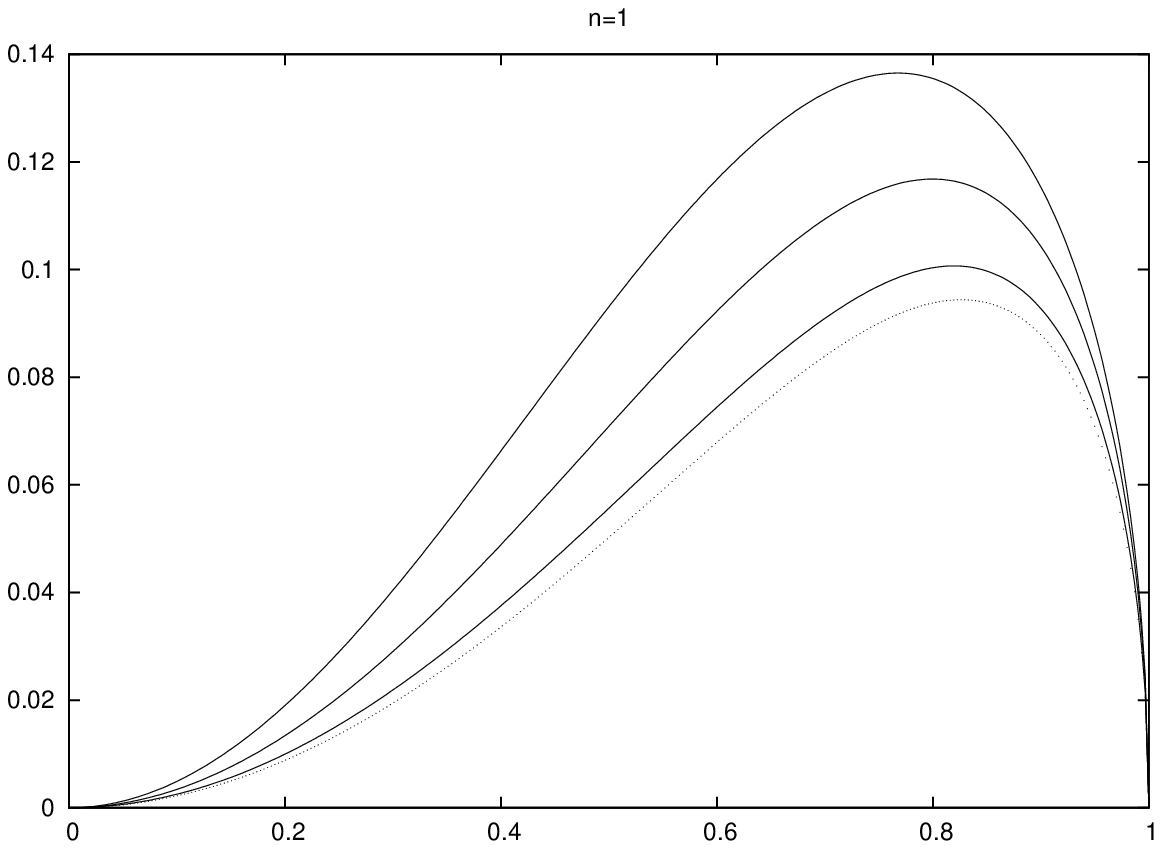}
\includegraphics[width=3.2in]{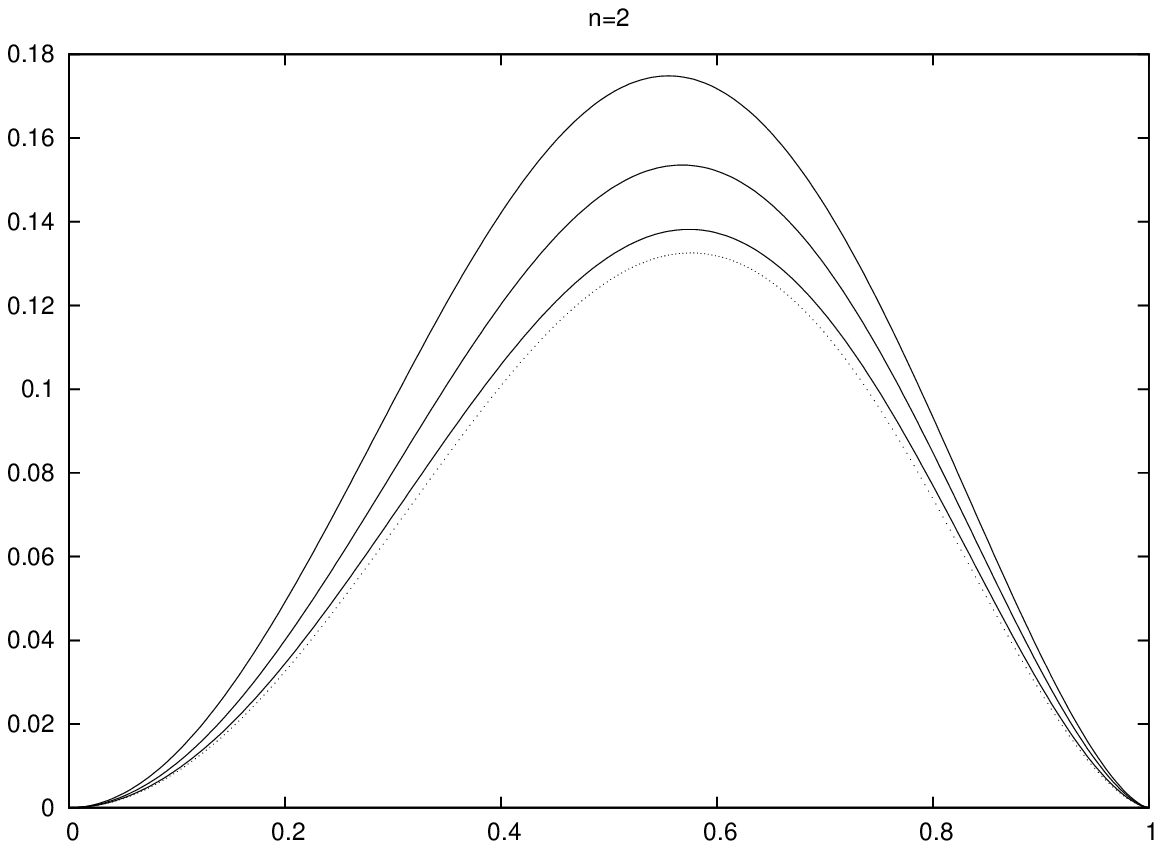}
\includegraphics[width=3.2in]{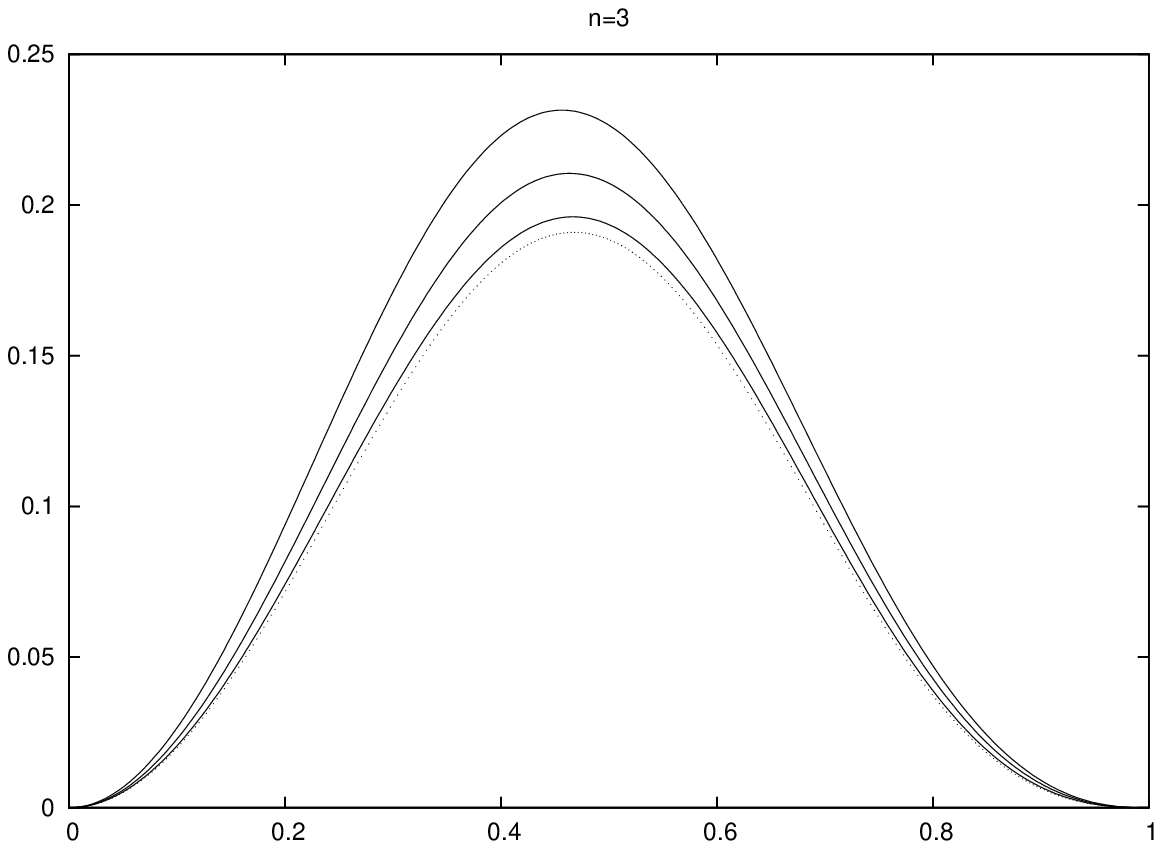}
\caption{\label{fig:pressure} Dimensionless azimuthal pressure ${\tilde
p}_n$ as a function of ${\tilde r}$ for the first three disk models  with
$n=1,2,3$. In each case, we plot ${\tilde p}_n(\tilde r)$ for $0 \leq
\tilde r \leq 1$ with $m = 0.10$ for different values of the parameter $\tilde
q$. First, we take $\tilde q =0.30$, the top curve in each plot, and then
$0.20,\;0.10$ and $\tilde q = 0$ (curve with dots), the vacuum case.}
\end{center}
\end{figure}

\begin{figure}
\begin{center}
\includegraphics[width=3.2in]{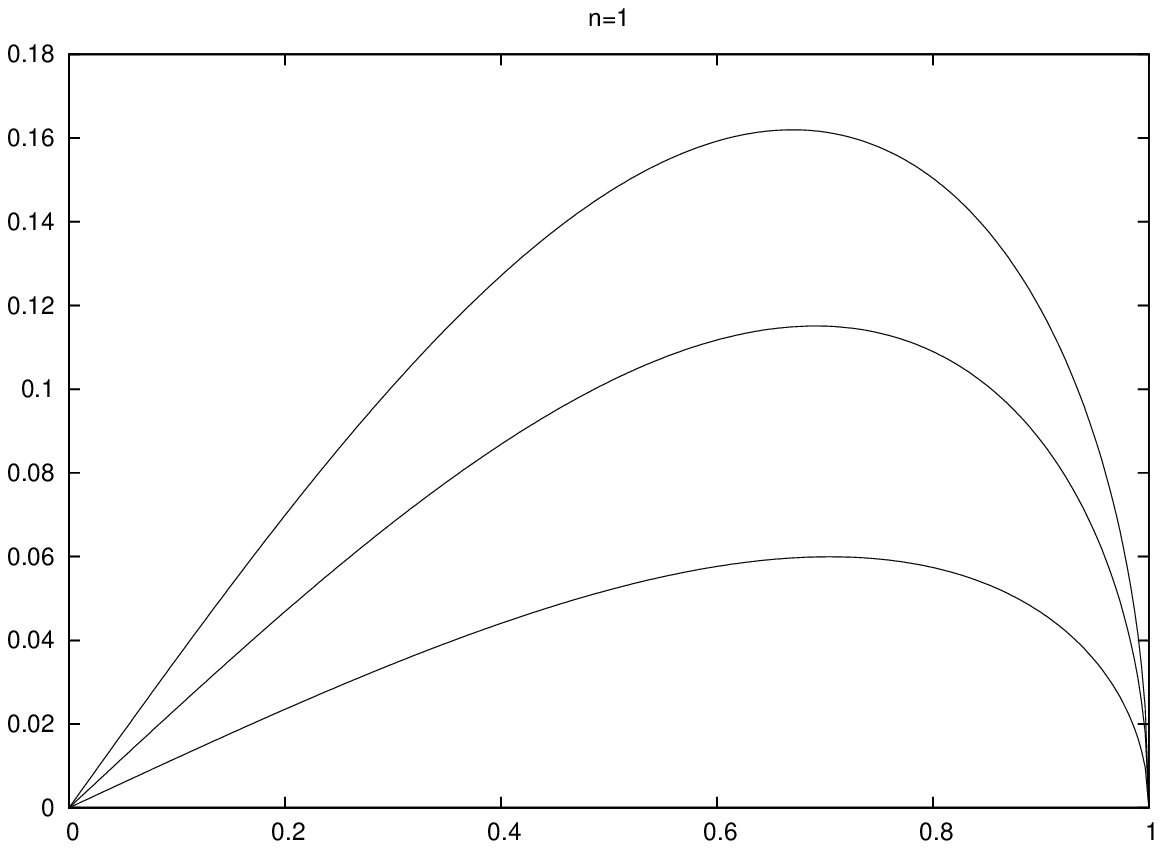}
\includegraphics[width=3.2in]{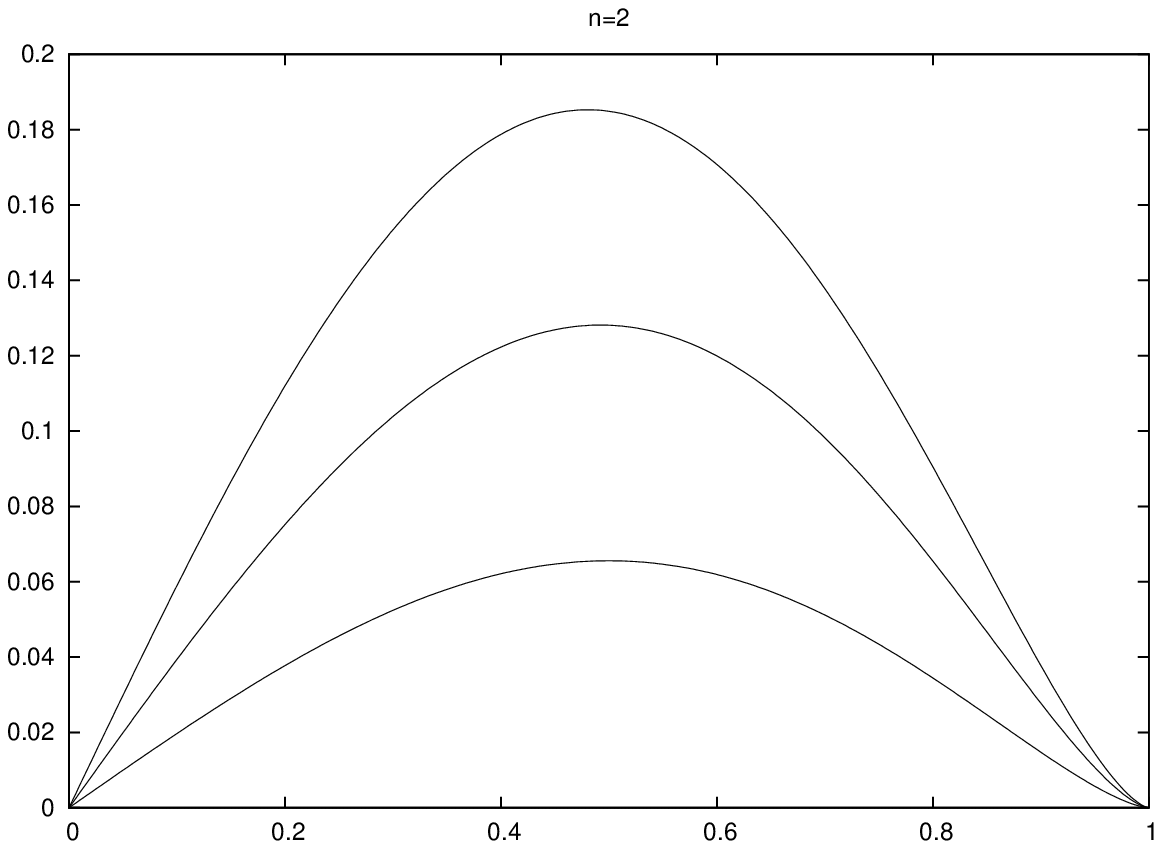} 
\includegraphics[width=3.2in]{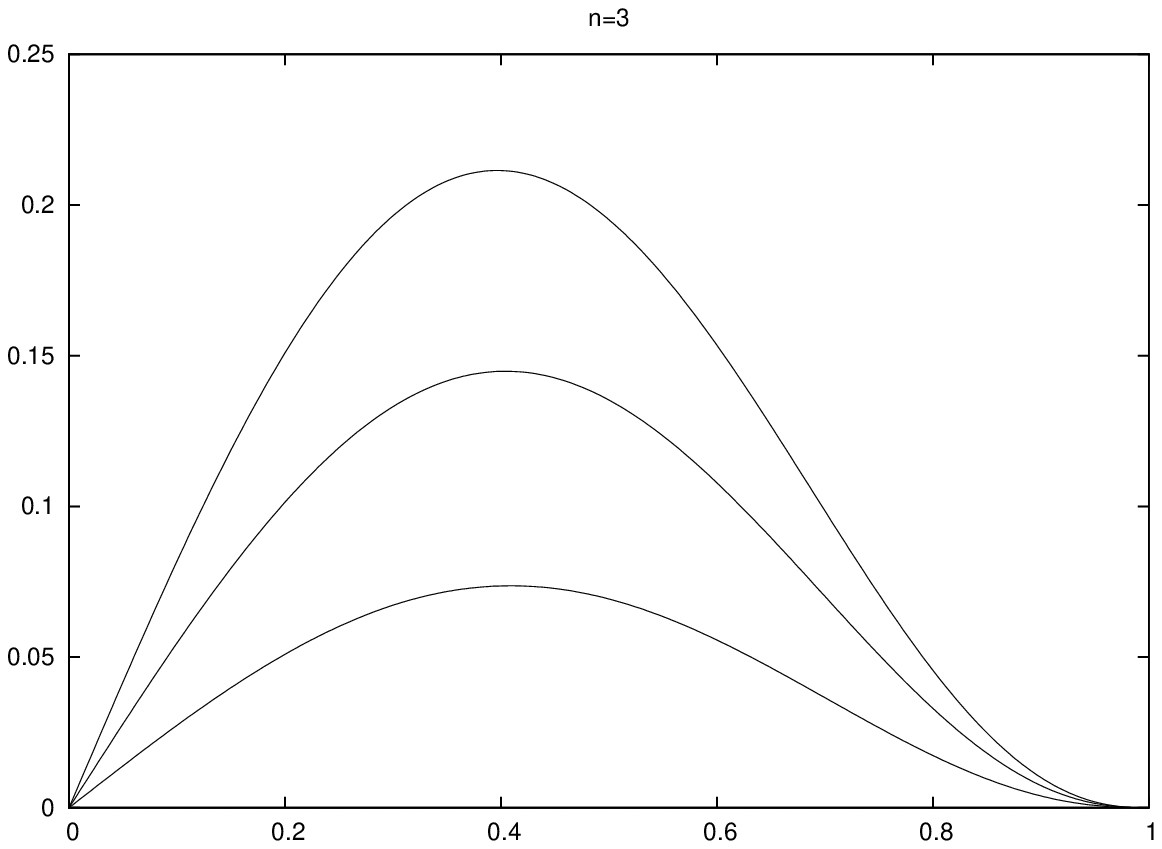}
\caption{\label{fig:current} Dimensionless surface current density ${\tilde
j}_n$ as a function of ${\tilde r}$ for the first three  disk models  with
$n=1,2,3$. In each case, we plot ${\tilde j}_n(\tilde r)$ for $0 \leq
\tilde r \leq 1$ with $m = 0.10$ for different values of the parameter $\tilde
q$. First, we take $\tilde q =0.30$, the top curve in each plot, and then $
0.20$ and $0.10$}
\end{center}
\end{figure}

The figure \ref{fig:pressure} shows the dimensionless azimuthal pressure,
$
p_n(r) = a \tilde p_n(\tilde r),
$
for the models corresponding to $n=1,\;2$ and $3$. As we see, in all the
cases, the pressure is  everywhere positive and rapidly increases as we move 
away from the center of the disk. It reaches a maximum and, later, it rapidly
decreases.  It can be observed that the presence of a magnetic field  increases
the pressure everywhere on the disk. Otherwise, the maximum increases and  moves
away from the disk edge as $n$ increases. It can be also observed that the
surface current density, represented in figure \ref{fig:current}, has a
behavior similar to that of the pressure. 

\begin{figure}
\begin{center}
\includegraphics[width=3.2in]{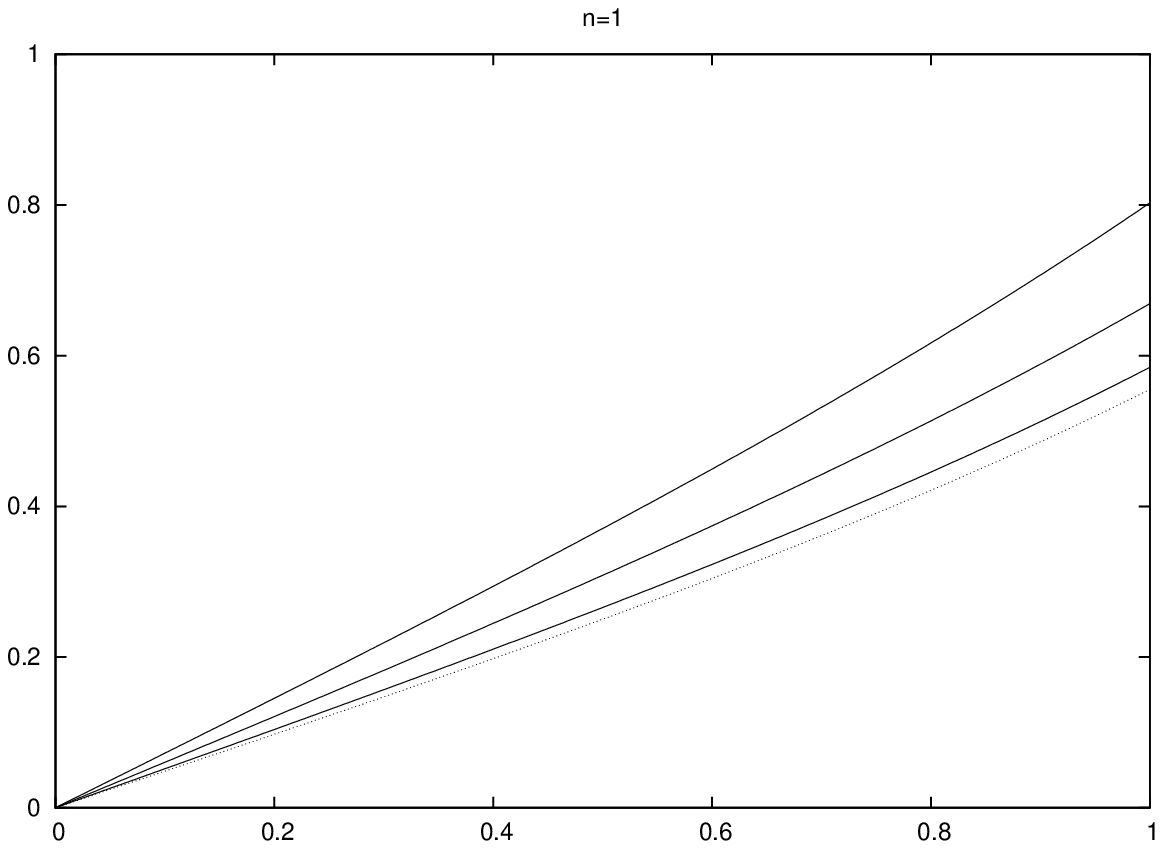}
\includegraphics[width=3.2in]{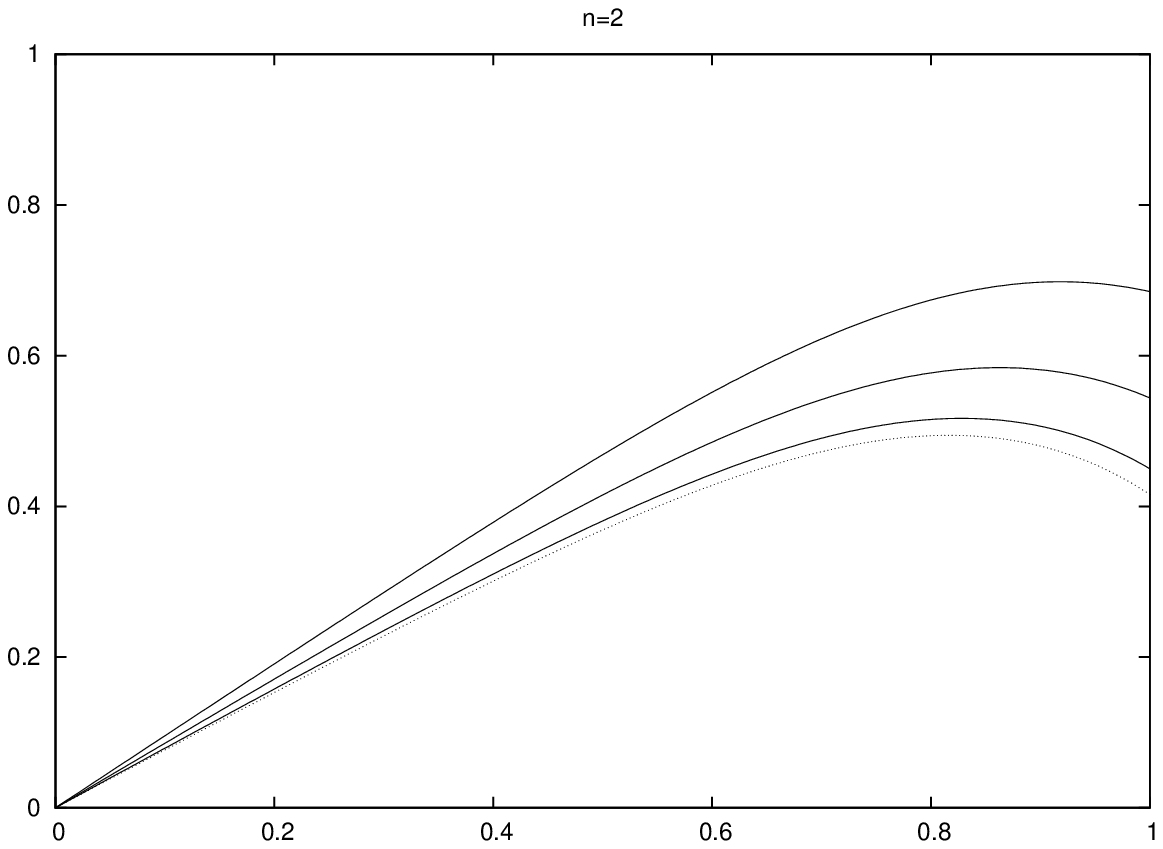} 
\includegraphics[width=3.2in]{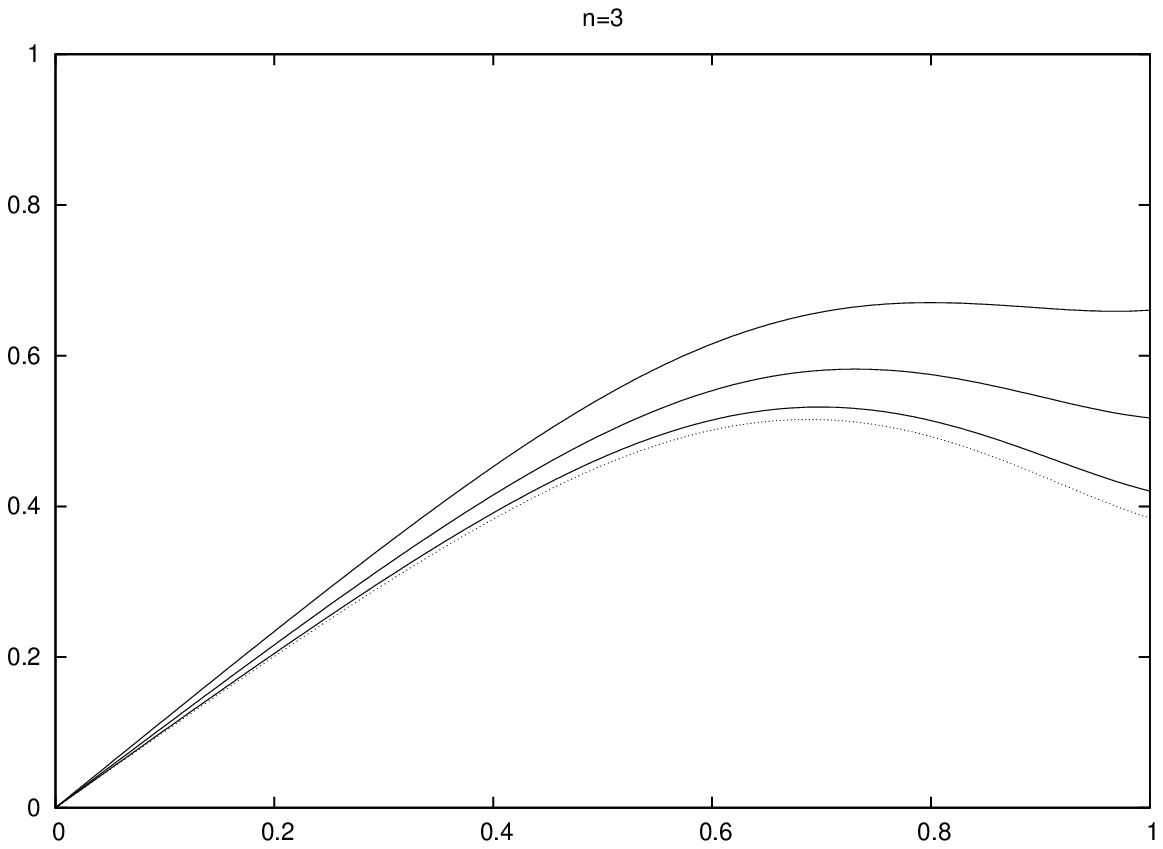}
\caption{\label{fig:velocity} Dimensionless circular velocity ${\tilde
U}_n$ as a function of ${\tilde r}$ for the three first disk models  with
$n=1,2,3$. In each case, we plot ${\tilde U}_n(\tilde r)$ for $0 \leq
\tilde r \leq 1$ with $m = 0.10$ for different values of the parameter $\tilde
q$. We first take $\tilde q =0.30$, the top curve in each plot, and then
$0.20,\;0.10$ and $\tilde q = 0$ (curve with dots), the vacuum case.}
\end{center}
\end{figure}

In order to  illustrate graphically the behavior of the circular velocities
or rotation curves, we introduce the dimensionless quantity
$
U_n(r) = a \tilde U_n(\tilde r).
$
In figure \ref{fig:velocity} we plot the dimensionless rotation curves for the
models corresponding to $n=1,\;2$ and $3$. The circular velocity corresponding
to $n=1$ is a monotonously increasing function of the radius. On the other hand,
for $n=2$ and $n=3$, the circular velocity increases from a value of zero at the
 center of the disks until it attains a maximum at a critical radius and then,
it decreases to acquire a finite value at the edge of the disk. It can also be
seen that the value of the critical radius increases as the values of $n$
decreases and $\tilde q$ increases. From which, we  can conclude that the
magnetic field decreases the circular velocity everywhere on the disk. We  study
these solutions with other values of the parameters as well, but, in all the
cases, we find  a similar behavior. We see that the magnitude of $m$ and $\tilde
q$ are limited uniquely by the condition that the magnitude of the velocities of
the disk cannot exceed the speed of light. The explicit calculation of the 
Kretschmann invariant $R_{abcd}R^{abcd}$ is too lengthy to be reproduced here.
A direct computation by the use of REDUCE shows that the first member of the
family of the magnetized Morgan-Morgan disk, $n=1$,
has a singularity at the rim of the disk where the Kretschmann invariant becomes
infinite although its mass density is finite everywhere, when all the $n >
1$ disks are regular. Obviously, this  property in the curvature is, of course,
inherited from the seed Morgan and Morgan metric disk (see \cite{S}). 

The solution of (\ref{eq:met3}) in terms of the seed solution  (Morgan and
Morgan metric disk, $\Phi_s = \phi$ ) can be written in the form 
\begin{eqnarray}
\Phi = \ln[e^{2\Phi_s} + q^2r^2] - \Phi_s \label{eq:hp}. 
\end{eqnarray} 
Therefore, when we take $r\rightarrow\infty$ clearly we see that, unlike the
Morgan and Morgan seed solution, generated spacetimes are not asymptotically
flat. Consequently, although the technique described here produces a new class
of solution of Einstein-Maxwell's equations and all the physical quantities
exhibit  an excellent behavior, the method offers  very little insight into the
physical implications of the metrics. The reason of this non-flatness  far away
from the disk should be that the magnetic field itself does not decay at
infinity as it appears to be easily expressible from  (\ref{eq:hp2}). On the
other hand, this still appears to be a viable solution of the Einstein-Maxwell
system but, apparently, corresponds to  magnetized disks immersed in a Melvin
magnetic universe.

\section{Concluding remarks}\label{sec:conc}

We have presented an infinite family of new exact solutions of the vacuum
Einstein-Maxwell equations for static and axially symmetric spacetimes. The
solutions describe an infinite family of magnetized finite thin  disks, the
magnetized version of the family of Morgan and Morgan relativistic thin
disks \cite{MM1}. The first member of the family of the magnetized Morgan-Morgan
disk, $n=1$, has a singularity at the rim of the disk where the Kretschmann
invariant becomes infinite, although its mass density is finite everywhere.
Whereas all the $n > 1$ disks are regular. Obviously, this  property in the
curvature is, of course, inherited from the seed Morgan and Morgan metric disk. 

Unlike the Morgan and Morgan seed solution, the generated spacetimes are not
asymptotically flat. Consequently, although the technique described here
produces a new class of solution of Einstein-Maxwell's equations and all the
physical quantities exhibit  an excellent behavior and the energy-momentum
tensor is in fully agreement with all the energy conditions, the method offers
very little insight into the physical implications of the metrics. However,
these are the first fully integrated exact solutions for this kind of magnetized
thin disk sources. They are important in so far  as they represent a new family
of exact solutions of the Einstein-Maxwell vacuum equations. Moreover, the
outlined method may serve as a guideline to find more physically acceptable
solutions in future works.

\section*{Appendix}

We pass on to the definition of  the Hodge $\boldsymbol{^{\star}}$  (star)
operation. Let's us $M$ a $m-$dimensional manifold endowed with a metric $g$.
The  Hodge $\boldsymbol{^{\star}}$  (star) operation is a  map: $\Omega^r(M)
\rightarrow \Omega^{r-m}(M) $ whose action, on the basis of the vector
$\Omega^r(M)$, is defined by \cite{MN} 
\begin{eqnarray*}
\boldsymbol{^{\star}}(\boldsymbol{dx}
^{\mu_1}\wedge\boldsymbol{dx}^{\mu_2}\wedge\ldots\wedge\boldsymbol{dx}^{\mu_r})
=\frac{\sqrt{|g|}}{(m-r)!}
\varepsilon^{{\mu_1}{\mu_2}\ldots{\mu_r}}
_{\qquad {\mu_{r+1}}\ldots{\mu_m}}
\boldsymbol{dx}^{\mu_{r+1}}\wedge\ldots\wedge
\boldsymbol{dx}^{\mu_m}.\nonumber
\end{eqnarray*}
For a $r-$form
\begin{eqnarray*}
\boldsymbol{A} = \frac{1}{r!}A_{{\mu_1}{\mu_2}\ldots{\mu_r}}\boldsymbol{dx}
^{\mu_1}\wedge\boldsymbol{dx}^{\mu_2}\wedge\ldots\wedge\boldsymbol{dx}^{\mu_r}
\in\Omega^r
\end{eqnarray*}
we have
\begin{eqnarray*}
\boldsymbol{^{\star} A} = \frac{\sqrt{|g|}}{r!(r-m)!}
A_{{\mu_1}{\mu_2}\ldots{\mu_r}}
\varepsilon^{{\mu_1}{\mu_2}\ldots{\mu_r}}
_{\qquad {\mu_{r+1}}\ldots{\mu_m}}
\boldsymbol{dx}^{\mu_{r+1}}\wedge\ldots
\wedge\boldsymbol{dx}^{\mu_m}\in\Omega^{m-r},
\end{eqnarray*}
where the totally anti-symetric tensor $\varepsilon$ is 
\begin{eqnarray*}
\varepsilon_{{\mu_1}{\mu_2}\ldots{\mu_m}} = \left\{ \begin{array}{ll}
+ 1\;\;\;\mbox{if}\;\;\;({\mu_1}{\mu_2}\ldots{\mu_m})\\\;\;\;
\mbox{is an even permutation of}\;\;\;(12\ldots m),\\
\nonumber\\
- 1\;\;\;\mbox{if}\;\;\;({\mu_1}{\mu_2}\ldots{\mu_m})\\\;\;\;
\mbox{is an odd permutation of}\;\;\;(12\ldots m),\\
\nonumber\\
\;\;\;0\;\;\;\mbox{otherwise},
\end{array} \right. \\
\end{eqnarray*}
and ``$\wedge$'' is the usual exterior product or wedge product.

\begin{acknowledgements}
A. C. G-P. wants to acknowledge financial support from COLCIENCIAS, Colombia.
\end{acknowledgements}


\end{document}